\documentclass[aps,twocolumn,showpacs,groupedaddress]{revtex4}

\newcommand{\Tr}{\text{Tr}}
\usepackage{graphicx}
\begin{document}
\title{Effect of inter-subsystem couplings on the evolution of composite systems }
\author{X. X. Yi, H. T. Cui, Y. H. Lin, and H. S. Song}
\affiliation{Department of physics, Dalian University of
Technology, Dalian 116024 China }
\date{\today}
\begin{abstract}
The effect of inter-subsystem coupling on the adiabaticity of
composite systems and that of its subsystems is investigated.
Similar to the adiabatic evolution defined for pure states,
non-transitional evolution for mixed states is introduced;
conditions for the non-transitional evolution are  derived and
discussed.  An example that describes two coupled qubits is
presented to detail the general presentation. The effects due to
non-adiabatic evolution on the geometric phase are also presented
and discussed.
\end{abstract}
 \maketitle

The study on the adiabaticity of quantal system may be traced back
to the mid 1980s, when Berry \cite{berry} conceived that a quantal
system in an eigenstate, adiabatically transport round a circuit
by varying parameters $\vec{R}$ in its Hamiltonian $H(\vec{R})$,
will acquire a geometric phase in addition to the familiar
dynamical phase factor. Since then geometric phase became an
interesting subject and has been extensively studied
\cite{shapere,thouless,sun} and generalized to non-adiabatic
evolution \cite{aharonov}, mixed states
\cite{uhlmann86,sjoqvist1,singh}, and open systems \cite{carollo}.
The geometric phase of a composite system in particular has
attracted a lot of attention for its possible applications  in
quantum information processing, where the whole set of universal
quantum gates are achieved based on the Abelian and/or non-Abelian
geometric operations \cite{zanardi1,jones,falci,duan,wang}. In
view of the geometric computation, the adiabaticity of the
composite system is of course an important issue, because it would
determine how well the system follows the loops. Nevertheless,
thorough studies aimed to address this issue, in particular for a
composite system with inter-subsystem coupling, are still few and
certainly not exhaustive \cite{yi}.

On the other hand, the geometric phase for mixed states is a new
subject and much remains to be understood. Uhlmann
\cite{uhlmann86}was the first to address this issue and later
Sj\"oqvist {\it et al.} formulated it from the viewpoint of
quantum interferometry  \cite{sjoqvist1}. This formulation is
available when the system undergoes  an unitary evolution. For
subsystems that compose a composite system with non-zero
inter-subsystem couplings, however, the evolution of each
subsystem is not unitary in general. This problem was explored in
a  recent paper\cite{yi} for a very rare situation when both the
composite system and its subsystems evolve adiabatically, but how
the subsystems may evolve while the composite system transport
adiabatically (or non-adiabatically) remains an open question.

In this paper, we will address these issues by investigating  the
adiabaticity of a composite system  that consists   of two coupled
spin-$\frac 1 2 $ subsystems or a pair of quantum bit. We analyze
the case where one of the spin-$\frac 1 2 $ is driven by a
precessing magnetic field, a case of relevance to Nuclear Magnetic
Resonance(NMR) quantum computation \cite{gershenfeld97} as well as
to the test of mixed state geometric phases \cite{du03}. We
calculate and analyze the effects of spin-spin couplings on the
adiabaticity of the composite system and its counterpart  of the
subsystems, four different kinds of time evolution are clarified
and illustrated, conditions for those evolutions   are presented
and discussed.

Let a composite system be govern by the Hamiltonian
\begin{equation}
H=H_1+H_2+H_{12}, \label{ha}
\end{equation}
where $H_i (i=1,2)$ denote the free Hamiltonian of subsystem $i$
and $H_{12}$ stands for the interaction between them. We suppose
that the Hamiltonian $H$ is changed by varying parameters
$\vec{R}=(X, Y,...)$ on which it depends. Then the excursion of
the system between times $t=0$ and $t=T$ can be pictured as
transport round a closed path $\vec{R}(t)$ in parameter space with
Hamiltonian $H(\vec{R}(t))$ and such that $\vec{R}(T)=\vec{R}(0)$.
At any instant, the natural basis consists of the eigenstates
$|\phi(\vec{R})\rangle$ of $H(\vec{R})$ for $\vec{R}=\vec{R}(t)$,
that satisfy $H(\vec{R})|\phi_n(\vec{R})\rangle={\cal
E}_n(\vec{R})|\phi_n(\vec{R})\rangle$, with energies ${\cal
E}_n(\vec{R})$. If $H(\vec{R})$ is altered slowly such that
\begin{equation}
|\frac{\langle \phi_n(\vec{R})|\frac{\partial}{\partial \vec{R}}
|\phi_m(\vec{R})\rangle\frac{d R}{dt}}{{\cal E}_n(\vec{R})-{\cal
E}_m(\vec{R})}|<<1, \label{adia1}
\end{equation}
it follows from the adiabatic theorem that at any instant the
system will be in an eigenstate of the instantaneous Hamiltonian.
In particular, if the Hamiltonian is returned to its original
form, the composite system will return to its original state,
apart from a phase factor. Eq.(\ref{adia1}) is the well known
condition for the adiabatic theorem to hold.

We next develop a generalization for the subsystems, going back to
the original adiabatic scenario in which the system returns to its
original state, but now taking mixed states into account  instead
of pure states. To this end, we first of all define {\it
non-transitional evolution} for  mixed states \cite {erik1}, this
definition is non-trivial in particular for subsystems that have
no  effective Hamiltonian available for it\cite{explain}. Let a
state $\rho(t)$ of the subsystem (say, subsystem 1) be written in
the diagonal form of $\rho(t)=\sum_i p_i(t)|E_i(t)\rangle\langle
E_i(t)|$, $\rho(t)$ depends on time via $\vec{R}(t)$ and we would
write the time-dependence of $\vec{R}(t)$ explicitly. It is clear
that $p_i(t)$ gives the probability of the subsystem in state
$|E_i(t)\rangle$. This form of writing is called the spectral
representation, while $p_i(t)$ denotes the eigenvalues and
$|E_i(t)\rangle$ the corresponding eigenvectors of $\rho(t)$. One
special case is that $p_i(t)$ (for any $i$) are time-independent,
this is a rare situation which implies no transitions among the
eigenstates of $\rho(t)$ when the composite system experiences
transport along the parameter loops. The subsystem in this state
with time-independent coefficients $p_i(t)$, i.e., $p_i(t)=p_i(0)$
independent of the varying parameters, is defined to undergo {\it
non-transitional evolution}, and the corresponding eigenstates
$|E_i(t)\rangle$ will be called non-transitional eigenstates.
Obviously the non-transitional evolution would return to the
adiabatic evolution when the states $|E_i(t)\rangle$ are the
eigenstates of the subsystem's Hamiltonian(if any available).
Moreover this definition is meaningful  even if there is no
Hamiltonian available for the subsystems,  a general situation for
coupled multi-particle systems. Thus the definition could find
broad use instead of the adiabatic evolution for pure states in
composite systems. Now we will drive a condition for the subsystem
to undergo this non-transitional evolution. For the composite
system govern by  Hamiltonian Eq.(\ref{ha}), a state
$|\psi(t)\rangle$ may be decomposed into Schmidt form
\begin{equation}
|\psi(t)\rangle=\sum_i\sqrt{p_i(t)}e^{-i\int_0^tH_{ii}(t^{'})
dt^{'}} |E_i(t)\rangle_1|e_i(t)\rangle_2, \label{adias}
\end{equation}
with $H_{ij}(t)$=$_1\langle E_i(t)|$ $ _2\langle
e_i(t)|H|e_j(t)\rangle_2|E_j(t)\rangle_1$ and
$|e_j(t)\rangle_2|E_j(t)\rangle_1\equiv|e_j(t)\rangle_2\otimes|E_j(t)\rangle_1$.
The reduced density matrix for the subsystem 1 follows
straightforwardly from Eq.(\ref{adias}) that
$\rho_1(t)=\Tr_2(|\psi(t)\rangle\langle
\psi(t)|)=\sum_ip_i(t)|E_i(t)\rangle\langle E_i(t)|.$  To find the
condition of the non-transitional evolution is now equivalent to
finding  conditions for $p_i(t)$ in Eq.(\ref{adias}) to be
time-independent. In units with $\hbar=1$, $|\psi(t)\rangle$
satisfies
\begin{equation}
 i\frac{\partial}{\partial t}|\psi(t)\rangle=H|\psi(t)\rangle,
\label{sch}
\end{equation}
where here and hereafter time-dependence are understood where not
written explicitly and $|E_i(t)\rangle$ ($|e_i(t)\rangle)$ denotes
states for subsystem 1 (2) where suffix omitted. We would like to
note that $|E_i(t)\rangle|e_i(t)\rangle$ are not the instantaneous
eigenstates of $H$ in general, so under the action of $H$
transitions among those states would occur. As you will see, the
condition for the non-transitional evolution would be equivalent
to negligible  ratios of the transition amplitude to the
respective energy spacing. The derivative equation for $p_i(t)$
follows from Eq.(\ref{sch}) that,
\begin{eqnarray}
& \ &i\dot{\sqrt{p_j}}+\sqrt{p_j}\langle E_j|\dot{E}_j\rangle
+\sqrt{p_j}\langle
e_j|\dot{e}_j\rangle\nonumber\\
&-&\sum_{k\neq j}\sqrt{p_k}
exp\{-i\int_0^t(H_{kk}-H_{jj})dt^{'}\}H_{jk}=0. \label{peq}
\end{eqnarray}
The simplest approximation is to neglect the off-diagonal elements
on the grounds  that
\begin{equation}
|\frac{H_{jk}}{H_{jj}-H_{kk}}|<<1, \label{con1}
\end{equation}
having this approximation, Eq.(\ref{peq}) yields
\begin{equation}
\sqrt{p_j(t)}=\sqrt{p_j(0)}e^{i(\gamma_{1j}+\gamma_{2j})},
\end{equation}
where $\gamma_{1j}=i\int_0^t\langle E_j(\tau)|\frac{\partial
}{\partial \tau}|E_j(\tau)\rangle d\tau$, and
$\gamma_{2j}=i\int_0^t\langle e_j(\tau)|\frac{\partial }{\partial
\tau}|e_j(\tau)\rangle d\tau$. We may rewrite $|E_j(\tau)\rangle$
$(|e_j(\tau)\rangle$) to be $e^{i\gamma_{j1}}|E_j(\tau)\rangle$
($e^{i\gamma_{j2}}|e_j(\tau)\rangle$) such that $p_j(t)=p_j(0)$.
Usually, $|E_j(t)\rangle|e_j(t)\rangle$ are not the instantaneous
eigenstates of the Hamiltonian $H$, so $H_{jk}$ represent the
transition amplitude between states $|E_j(t)\rangle|e_j(t)\rangle$
and $|E_k(t)\rangle|e_k(t)\rangle$. Condition Eq.(\ref{con1})
indicates that for non-transitional evolutions, transitions
induced by $H$ among $|E_i(t)\rangle|e_i(t)\rangle$ should be
small with respect to the energy spacing between the two.
Conditions (\ref{con1}) and (\ref{adia1}) together imply that
there might be four different kinds  of evolution for the
composite system, as follows. (a){\it Adiabatic evolution}: The
composite system undergoes an adiabatic evolution while its
subsystems follow non-transitional evolutions, this means that the
composite system would follow one of its instantaneous eigenstates
while its subsystem evolve along the non-transitional eigenstates.
(b){\it Quasi-adiabatic evolution 1}: The composite system
undergoes an adiabatic evolution while its subsystems do not;
(c){\it Quasi-adiabatic evolution 2}: The composite system evolve
non-adiabatically while its subsystems follow non-transitional
evolutions. (d){\it Non-adiabatic evolution}: The composite system
evolve non-adiabatically, while its subsystems undergo out of
non-transitional evolutions.

It is worthwhile to mention that the condition/criterion for
non-transitional evolution given by Eq.(\ref{con1}) is also valid
for the  system with a mixed state under  unitary evolutions, this
can be understood as follows. Write the state (density matrix) of
the composite system in a spectral representation
\begin{equation}
\rho(t)=\sum_{\alpha}\lambda_{\alpha}|\psi_{\alpha}(t)\rangle\langle\psi_{\alpha}(t)|,
\end{equation}
$|\psi_{\alpha}(t)\rangle$ would obey the Schr\"odinger equation
and has the same form of  decomposition as Eq.(\ref{adias}) but
with $\sqrt{p_j^{\alpha}(t)}$ instead of $\sqrt{p_j(t)}$. The von
Neumann equation $i\hbar\frac{\partial\rho}{\partial t}=[H,\rho]$
then yields the same equation as  in Eq.(\ref{peq}) but with
$\sqrt{p_j}=\sqrt{\sum_{\alpha}p_j^{\alpha}\lambda_{\alpha}}$.
$\sum p_j^{\alpha}\lambda_{\alpha}$ represents the population of
the subsystem 1 in state $|E_j(t)\rangle$, non-transitional
evolution that requires $p_j$ constant would result in the same
condition as in Eq.(\ref{con1}), indeed for a closed system  the
transitions among the eigenstates of the subsystem's density
matrix are only induced by the system Hamiltonian, for  open
systems, however, this is not the case, since the bath which
couples to the whole system may lead to population transfer among
those states. To this respect, in addition to the condition
Eq.(\ref{con1}), criteria
\begin{equation}
|\frac{\langle E_k|\langle
e_k|\Gamma_{\alpha}|e_j\rangle|E_j\rangle}{H_{jj}-H_{kk}}|<<1
\end{equation}
is required to be satisfied in order to ensure the
non-transitional evolutions for  open systems, where
$\Gamma_{\alpha}$ represent the agents (operators) of the whole
system coupled to a bath. This is the reason why the
non-transitional condition should be different from each other for
an open system and for a closed system, and this is also an
explanation for closed systems, regardless of in a pure state or a
mixed state, hold the same criteria for non-transitional
evolutions.

As an  example, we consider two qubits $\vec{S}_k (k=1,2)$ as
represented by a pair of spin-$\frac 1 2 $ particles, coupled
through a uniaxial exchange interaction in the z-direction. One of
the qubits (say, qubit 1) is driven by a time-dependent magnetic
field $\vec{ B}(t) =B_0 \hat{{\bf n}}(t)$ with the unit vector
$\hat{{\bf
n}}=(\sin\theta\cos\phi,\sin\theta\sin\phi,\cos\theta)$, the
Hamiltonian of this system reads ($\hbar=1$) \cite{abragam}
\begin{equation}
H(t) = 4J S_1^z \otimes S_2^z + \mu \vec{ B}(t) \cdot \vec{ S}_1
\label{hatq}
\end{equation}
with the exchange interaction constant $J$ and the gyromagnetic
ratio $\mu$. This Hamiltonian is of relevance to NMR experiment
where Carbon-13 labelled chloroform in $d_6$ acetone may be used
as the sample. The single $^{13}C$ nucleus and the $^1H$ nucleus
play the role of the two spin-$\frac 1 2 $ particles; the
spin-spin coupling constant in this case is $4J\simeq (2\pi)214.5
\mbox{Hz}$. The individual addressing can be realized in NMR by
using spins of nuclei of different isotopes, such as those of
different species of atoms, as the subsystems, these spins usually
have precession frequencies that differ from the other by many
MHz, a resonant magnetic field for one spin then has little
effects on the others. For more detail, we refer the reader to
Ref.\cite{cory}.

The instantaneous eigenstates and the corresponding eigenvalues
(in units of $\frac 1 2 \mu B_0$) can be written as
\begin{widetext}
\begin{eqnarray}
|\phi_{1,2}(t)\rangle&=&\frac{1}{\sqrt{M_{1,2}}}(\sin\theta
e^{-i\phi}|\downarrow\uparrow\rangle +(g+\cos\theta+{\cal
E}_{1,2})|\uparrow\uparrow\rangle),\nonumber\\
|\phi_{3,4}(t)\rangle&=&\frac{1}{\sqrt{M_{3,4}}}(\sin\theta
e^{-i\phi}|\downarrow\downarrow\rangle +(\cos\theta+{\cal
E}_{3,4}-g)|\uparrow\downarrow\rangle),\label{eigenf}
\end{eqnarray}
\end{widetext}
and
\begin{eqnarray}
{\cal E}_{1,2}&=&\pm\sqrt{(g^2+1)+ 2 g\cos\theta},\nonumber\\
{\cal E}_{3,4}&=&\pm\sqrt{(g^2+1)- 2 g\cos\theta}, \label{eigenv}
\end{eqnarray}
where $g=\frac{2J}{\mu B_0}$ denotes the rescaled exchange
interaction constant, $M_j$ the renormalization constant and
$|\uparrow \downarrow\rangle=|\uparrow \rangle_1\otimes
|\downarrow\rangle_2$ and the others likely. For a simplest case
where the exchange interaction $g=0$, the eigenvalues are reduced
to ${\cal E}_{\pm}=\pm 1 $, with corresponding instantaneous
eigenstates $|\phi_{+}(t)\rangle= (\cos\frac{\theta}{2}$ $
|\uparrow \rangle
+\sin\frac{\theta}{2}e^{-i\phi})\otimes|\uparrow\rangle$ (or
$\otimes |\downarrow\rangle$),  $|\phi_{-}(t)\rangle=
(-\sin\frac{\theta}{2}$ $ |\uparrow \rangle
+\cos\frac{\theta}{2}e^{-i\phi})\otimes|\uparrow\rangle$ (or
$\otimes |\downarrow\rangle$).   Suppose the external magnetic
fields precess with time-independent azimuthal angles $\theta$ and
a constant precessing frequency $\omega$, i.e., $\phi=\omega t$,
the adiabatic evolution for the composite system requires $
|\langle\phi_+|\dot{\phi}_-\rangle/({\cal E}_+-{\cal E}_-)|<<1,$
that is (in units of $\mu B_0/2 $)
$\omega<<|\frac{4}{\sin\theta}|.$  Obviously, in this limit the
two spin-$\frac 1 2 $ particles remain uncoupled and the adiabatic
condition is exactly the one for the driven particles 1.  From the
aspect of non-transitional evolution, no constraints on $\omega$
could be made,  because of there is no coupling between the two
qubits and each qubit would remain in pure states if the initial
states are pure. So, for a composite system without
inter-subsystem couplings, the evolutions would fall in regime (a)
or (c), i.e., the composite system might undergo an adiabatic or
non-adiabatic evolution, while its subsystems evolve along the
non-transitional states certainly.

Now we turn to study the case with inter-subsystem couplings. For
the composite system, to make the adiabatic theorem valid, it
should be satisfied that $(i,j =1,...,4,i\neq j)$
\begin{widetext}
\begin{eqnarray}
\Gamma_{ij} &\equiv&
|\frac{\langle\phi_i|\dot{\phi}_j\rangle}{{\cal E}_i-{\cal E}_j}|
=\frac{1}{\sqrt{M_iM_j}}|\frac{\omega\sin^2\theta }{{\cal
E}_i-{\cal E}_j}|<<1, i,j=1,2 \ \ \mbox{or} \ \ \ i,j=3,4,
\nonumber\\
\Gamma_{ij}&=&0, \mbox{others}.
\end{eqnarray}
\end{widetext}
This condition follows straightforwardly from Eq.(\ref{adia1}) by
assuming the azimuthal angle $\theta$ time-independent and
$\phi=\omega t$. Clearly, the  eigenenergies ${\cal E}_i$ and the
renormalization constant $M_i$ are independent of $\omega$, so
$\Gamma_{ij}$ increase linearly with $\omega$, i.e., slowly
precessing magnetic fields would benefit the adiabatic evolution.
\begin{figure}
\includegraphics*[width=0.95\columnwidth,
height=0.6\columnwidth]{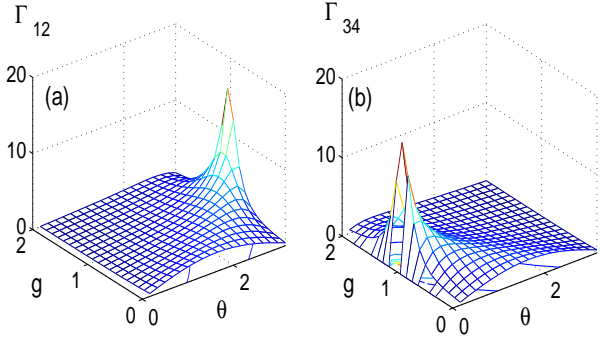} \caption{ (Color online)
Selected results for $\Gamma_{ij}$ as a function of the azimuthal
angle $\theta$[Arc] and the rescaled coupling constant $g$ (in
units of $\mu B_0/2 $). $\omega=10$(in units of $\mu B_0/2 $) was
chosen for this plot. Figures (a),(b) are for different $i$ and
$j$; (a) $\Gamma_{12}$, (b) $\Gamma_{34}$. } \label{fig1}
\end{figure}
The dependence of $\Gamma_{ij}$ $(i,j=1,2 \ \ \mbox {or} \ \ 3,4)$
on $g$ and $\theta$ was illustrated in figure 1. A common feature
of these figures is that $\Gamma_{ij}\rightarrow 0$ with the
rescaled coupling constant $g\rightarrow \infty$. This limit
corresponds to the case when the second term in the Hamiltonian
Eq.(\ref{hatq}) can be ignored. Physically, the inter-subsystem
coupling increase the energy spacing between any two instantaneous
eigenvalues, this makes the population transfer between the
respective instantaneous eigenstates more difficult, and equally
the adiabatic evolutions easier.
 $\Gamma_{12}$,  and $\Gamma_{34}$  behave as a non-monotonic function of $g$ as
 figures 1-(a), and (b) show, they increase to a maximum value
 for a disaster coupling and then towards to zero for a
 sufficiently large coupling.  Two singular points can be found in
 figure 1, for $\Gamma_{12}$ the point is located at $g=1$ and
 $\theta=\pi$, while for $\Gamma_{34}$, $g=1$ and $\theta=0$. This
 singularity results form the degeneracy of the corresponding
 eigenstates, and hence the adiabatic condition is difficult to
 meet at these points.

These transitions induced by non-adiabaticity among the
instantaneous eigenstates of the Hamiltonian would affect the
geometric phase of the composite system,  the geometric phase in
this case may be calculated by removing the accumulation of these
dynamical phases from the total phase, i.e.,
\begin{equation}
\phi_g=arg\langle\Psi(0)|\Psi(T)\rangle+i\int_0^T
dt\langle\Psi(t)|\dot{\Psi}(t)\rangle,\label{gp}
\end{equation}
where $|\Psi(t)\rangle$ represents the state of the composite
system at instance $t$, and can be written as
\begin{equation}
|\Psi(t)\rangle=\sum_{j=1}^4c_j(t)e^{-i\int_0^t{\cal E}_j(\tau)
d\tau }|\phi_j(t)\rangle,
\end{equation}
with $|\phi_j(t)\rangle$ the instantaneous eigenvector and ${\cal
E}_j$ the corresponding eigenvalue. The Schr\"odinger equation $
i\hbar \frac{\partial}{\partial
t}|\Psi(t)\rangle=H|\Psi(t)\rangle$ yields (with
$\Omega_{jk}(t)=\int_0^t{\cal E}_k(\tau)-{\cal E}_j(\tau) d\tau$)
\begin{eqnarray}
 & \ & \dot{c}_j(t)+\langle\phi_j(t)|\dot{\phi}_j(t)\rangle
 c_j(t)\nonumber\\
 &=&-\sum_{k\neq j}e^{i\Omega_{kj}(t)}\langle
 \phi_j(t)|\dot{\phi}_k(t)\rangle c_k(t),
 \end{eqnarray}
 the Berry's phase is just a follow up of this equation by
 ignoring its right-hand side, further consideration would treat
 the terms in the right-hand side as perturbations. Up to the
 first order in
 $|\frac{\langle\phi_j(t)|\dot{\phi}_k(t)\rangle}{{\cal E}_k-{\cal
 E}_j}|$, the usual perturbation theory give( assuming
 $c_n(t=0)=1$)
 \begin{eqnarray}
c_n(t)&\simeq& e^{i\gamma_n(t)},\nonumber\\
c_j(t)&\simeq&
\frac{e^{i\Omega_{kj}(t)+i\gamma_k(t)}\langle\phi_j(t)|\dot{\phi}_k(t)\rangle}{{\cal
E}_j-{\cal E}_k}, j\neq n \label{cc}
 \end{eqnarray}
 with
 $\gamma_p(t)=i\int_0^t\langle\phi_p(\tau)|\dot{\phi}_p(\tau)\rangle
 d\tau$, the Berry phase pertained to instantaneous eigenstate
 $|\phi_p(t)\rangle$.
 Eq.(\ref{gp}) and (\ref{cc}) together yield the geometric phase
 \begin{equation}
 \phi_g=arg[e^{i\gamma_n(T)}+\sum_{m\neq n}c_m(T)]. \label{bpa1}
 \end{equation}
To get this expression, all terms equal or smaller than
$\Gamma_{mn}^2$ have been ignored. It is clear that the main
correction caused by  the non-adiabatic evolution to the geometric
phase come from terms $\sum_{m\neq n} c_m(T)$, that may be
expressed up to first order in $\Gamma_{mn}$ as
\begin{eqnarray}
\phi_g&=&\gamma_n(T)+\sum_{m\neq
n}\Gamma_{mn}[\Omega_{nm}(T)+\gamma_m(T)\nonumber\\
&+&arg\langle\phi_n(T)|\dot{\phi}_m(T)\rangle],
\end{eqnarray}
with $\Gamma_{mn}\rightarrow 0$ the geometric phases approach the
Berry phases $\gamma_n(T)$ as expected. We may divide the
correction due to the non-adiabatic evolution to the geometric
phase into two kinds, the first is the population transfer among
the instantaneous eigenstates, this contribution appears in the
correction as $\sum_{m\neq
n}\Gamma_{mn}(\Omega_{nm}(T)+arg\langle\phi_n(T)|\dot{\phi}_m(T)\rangle)$,
while the second that exhibit the geometric feature of the
transited eigenstates scales as $\sum_{m\neq
n}\Gamma_{mn}\gamma_m(T)$. We would like to note that only
$\Gamma_{12}$ and $\Gamma_{34}$ are not zero in the example Eq.
(\ref{hatq}), but the representation Eq. (\ref{bpa1}) is quite
general for quantum systems.

 It is not difficult to show from Eq.(\ref{eigenf}) that the
 transport of the subsystems is always non-transitional in this
 situation. For example,
suppose the composite system undergo an adiabatic evolution in the
instantaneous eigenstate $|\phi_n(t)\rangle (n=1, 2)$, the reduced
density matrix of the subsystem 1 reads
\begin{widetext}
\begin{eqnarray}
\rho_1(t)&=&\frac {1}{M_n}\left (\matrix{ (g+\cos\theta+{\cal
E}_n)^2 & \sin\theta(g+{\cal E}_n+\cos\theta)e^{i\phi} \cr
\sin\theta(g+{\cal E}_n+\cos\theta)e^{-i\phi} & \sin^2\theta }
\right )\equiv \left ( \matrix{\rho_{11} & \rho_{12} \cr
\rho_{12}^* &
\rho_{22}}\right ) \nonumber\\
&=& \rho_+|\rho_+\rangle\langle
\rho_+|+\rho_-|\rho_-\rangle\langle \rho_-| \label{con2}
\end{eqnarray}
\end{widetext}
with $\rho_{+,-}=0,1$. Clearly, $\rho_{\pm}$ are time-independent.
This point will be changed  when  the spin-spin coupling takes the
form of $(J\sigma_1^+\sigma_2^-+h.c.)$, the instantaneous
eigenstates in this situation are entangled states of the two
particles, the diagonal elements of the reduced density matrix
$\rho_1$ is $\theta$-dependent as Ref.\cite{yi} shown, and the
slowly varying $\theta$ can make the composite system an adiabatic
evolution meanwhile make the subsystems out of non-transitional
evolution.

Some remarks on the non-transitional evolutions are now in order.
For subsystems with an available Hamiltonian, the concept of
non-transitional evolution covers the concept of adiabatic
evolution, this can be understood as follows. Write an initial
state (generally mixed) of the subsystem in the natural basis (the
instantaneous eigenstates of the subsystem's Hamiltonian)
$\rho_1(0)=\sum_{i,j}\rho_{ij}|\Phi_i(0)\rangle\langle
\Phi_j(0)|,(i,j=1,...,4),$ adiabatic evolution yields
$\rho_1(t)=\sum_{i,j}\rho_{ij}|\Phi_i(t)\rangle\langle
\Phi_j(t)|,(i,j=1,...,4).$ Diagonalizing $\rho(t)$, we can get
non-negative and time-independent eigenvalues since
$\rho_{ij}=\rho_{ji}^*$, this means that the usual adiabatic
evolution must fall in the regime of the non-transitional
evolution, but the inverse could not be proven correct.

In conclusion, the evolution of composite systems and its
subsystems has been studied. By the definition of non-transitional
evolution, four different kinds of evolution were identified and
illustrated via the coupled two-qubit system. The non-transitional
evolution would find its use in formulating evolution of composite
systems, in particular for subsystems that have no Hamiltonian
available.

\vskip 0.3 cm We acknowledge financial support from EYTP of M.O.E,
and NSF of China, Project No. 10305002.


\begin{thebibliography}{99}
\bibitem{explain} This means that we could not find an effective
Hamiltonian that governs  the evolution of the subsystems, a
simple example is when $H_{12}$ in Eq.(\ref{ha}) is not zero, and
$[H_1,H_{12}]\neq 0$ (the subsystem 1 has no an effective
Hamiltonian), and $[H_2,H_{12}]\neq 0$ (the subsystem 2 has no an
effective Hamiltonian).
\end{thebibliography}
\end{document}